\documentclass[aps,prc,superscriptaddress,twocolumn,nofootinbib,amsmath,amssymb,showpacs,showkeys,floatfix]{revtex4-1}
\usepackage{graphicx}
\usepackage{feynmf}
\usepackage[caption=false]{subfig}

\begin{document}




\title{First Observation of the $\Lambda(1405)$ Line Shape in Electroproduction}

%
%
%
%
%
%
%
%

\newcommand*{\CMU}{Carnegie Mellon University, Pittsburgh, Pennsylvania 15213}
\newcommand*{\CMUindex}{1}
\affiliation{\CMU}
\newcommand*{\ANL}{Argonne National Laboratory, Argonne, Illinois 60439}
\newcommand*{\ANLindex}{2}
\affiliation{\ANL}
\newcommand*{\ASU}{Arizona State University, Tempe, Arizona 85287-1504}
\newcommand*{\ASUindex}{3}
\affiliation{\ASU}
\newcommand*{\CSUDH}{California State University, Dominguez Hills, Carson, CA 90747}
\newcommand*{\CSUDHindex}{4}
\affiliation{\CSUDH}
\newcommand*{\CANISIUS}{Canisius College, Buffalo, NY 14208}
\newcommand*{\CANISIUSindex}{4}
\affiliation{\CANISIUS}
\newcommand*{\CUA}{Catholic University of America, Washington, D.C. 20064}
\newcommand*{\CUAindex}{5}
\affiliation{\CUA}
\newcommand*{\SACLAY}{CEA, Centre de Saclay, Irfu/Service de Physique Nucl\'eaire, 91191 Gif-sur-Yvette, France}
\newcommand*{\SACLAYindex}{6}
\affiliation{\SACLAY}
\newcommand*{\CNU}{Christopher Newport University, Newport News, Virginia 23606}
\newcommand*{\CNUindex}{7}
\affiliation{\CNU}
\newcommand*{\UCONN}{University of Connecticut, Storrs, Connecticut 06269}
\newcommand*{\UCONNindex}{8}
\affiliation{\UCONN}
\newcommand*{\EDINBURGH}{Edinburgh University, Edinburgh EH9 3JZ, United Kingdom}
\newcommand*{\EDINBURGHindex}{9}
\affiliation{\EDINBURGH}
\newcommand*{\FU}{Fairfield University, Fairfield CT 06824}
\newcommand*{\FUindex}{10}
\affiliation{\FU}
\newcommand*{\FIU}{Florida International University, Miami, Florida 33199}
\newcommand*{\FIUindex}{11}
\affiliation{\FIU}
\newcommand*{\FSU}{Florida State University, Tallahassee, Florida 32306}
\newcommand*{\FSUindex}{12}
\affiliation{\FSU}
\newcommand*{\Genova}{Universit$\grave{a}$ di Genova, 16146 Genova, Italy}
\newcommand*{\Genovaindex}{13}
\affiliation{\Genova}
\newcommand*{\GWUI}{The George Washington University, Washington, DC 20052}
\newcommand*{\GWUIindex}{14}
\affiliation{\GWUI}
\newcommand*{\ISU}{Idaho State University, Pocatello, Idaho 83209}
\newcommand*{\ISUindex}{15}
\affiliation{\ISU}
\newcommand*{\INFNFE}{INFN, Sezione di Ferrara, 44100 Ferrara, Italy}
\newcommand*{\INFNFEindex}{16}
\affiliation{\INFNFE}
\newcommand*{\INFNFR}{INFN, Laboratori Nazionali di Frascati, 00044 Frascati, Italy}
\newcommand*{\INFNFRindex}{17}
\affiliation{\INFNFR}
\newcommand*{\INFNGE}{INFN, Sezione di Genova, 16146 Genova, Italy}
\newcommand*{\INFNGEindex}{18}
\affiliation{\INFNGE}
\newcommand*{\INFNRO}{INFN, Sezione di Roma Tor Vergata, 00133 Rome, Italy}
\newcommand*{\INFNROindex}{19}
\affiliation{\INFNRO}
\newcommand*{\ORSAY}{Institut de Physique Nucl\'eaire ORSAY, Orsay, France}
\newcommand*{\ORSAYindex}{20}
\affiliation{\ORSAY}
\newcommand*{\ITEP}{Institute of Theoretical and Experimental Physics, Moscow, 117259, Russia}
\newcommand*{\ITEPindex}{21}
\affiliation{\ITEP}
\newcommand*{\JMU}{James Madison University, Harrisonburg, Virginia 22807}
\newcommand*{\JMUindex}{22}
\affiliation{\JMU}
\newcommand*{\KNU}{Kyungpook National University, Daegu 702-701, Republic of Korea}
\newcommand*{\KNUindex}{23}
\affiliation{\KNU}
\newcommand*{\LPSC}{LPSC, Universite Joseph Fourier, CNRS/IN2P3, INPG, Grenoble, France
}
\newcommand*{\LPSCindex}{24}
\affiliation{\LPSC}
\newcommand*{\UNH}{University of New Hampshire, Durham, New Hampshire 03824-3568}
\newcommand*{\UNHindex}{25}
\affiliation{\UNH}
\newcommand*{\NSU}{Norfolk State University, Norfolk, Virginia 23504}
\newcommand*{\NSUindex}{26}
\affiliation{\NSU}
\newcommand*{\OHIOU}{Ohio University, Athens, Ohio  45701}
\newcommand*{\OHIOUindex}{27}
\affiliation{\OHIOU}
\newcommand*{\ODU}{Old Dominion University, Norfolk, Virginia 23529}
\newcommand*{\ODUindex}{28}
\affiliation{\ODU}
\newcommand*{\RPI}{Rensselaer Polytechnic Institute, Troy, New York 12180-3590}
\newcommand*{\RPIindex}{29}
\affiliation{\RPI}
\newcommand*{\URICH}{University of Richmond, Richmond, Virginia 23173}
\newcommand*{\URICHindex}{30}
\affiliation{\URICH}
\newcommand*{\ROMAII}{Universita' di Roma Tor Vergata, 00133 Rome Italy}
\newcommand*{\ROMAIIindex}{31}
\affiliation{\ROMAII}
\newcommand*{\MSU}{Skobeltsyn Nuclear Physics Institute, 119899 Moscow, Russia}
\newcommand*{\MSUindex}{32}
\affiliation{\MSU}
\newcommand*{\SCAROLINA}{University of South Carolina, Columbia, South Carolina 29208}
\newcommand*{\SCAROLINAindex}{33}
\affiliation{\SCAROLINA}
\newcommand*{\JLAB}{Thomas Jefferson National Accelerator Facility, Newport News, Virginia 23606}
\newcommand*{\JLABindex}{34}
\affiliation{\JLAB}
\newcommand*{\UTFSM}{Universidad T\'{e}cnica Federico Santa Mar\'{i}a, Casilla 110-V Valpara\'{i}so, Chile}
\newcommand*{\UTFSMindex}{35}
\affiliation{\UTFSM}
\newcommand*{\GLASGOW}{University of Glasgow, Glasgow G12 8QQ, United Kingdom}
\newcommand*{\GLASGOWindex}{36}
\affiliation{\GLASGOW}
\newcommand*{\VIRGINIA}{University of Virginia, Charlottesville, Virginia 22901}
\newcommand*{\VIRGINIAindex}{37}
\affiliation{\VIRGINIA}
\newcommand*{\WM}{College of William and Mary, Williamsburg, Virginia 23187-8795}
\newcommand*{\WMindex}{38}
\affiliation{\WM}
\newcommand*{\YEREVAN}{Yerevan Physics Institute, 375036 Yerevan, Armenia}
\newcommand*{\YEREVANindex}{39}
\affiliation{\YEREVAN}
 
\newcommand*{\NOWIOWA}{University of Iowa, Iowa City, IA 52242}
\newcommand*{\NOWINDIANA}{Indiana University, Bloomington, Indiana 47405}
\newcommand*{\NOWSIENA}{Siena College, Loudonville, NY 12211}
\newcommand*{\NOWMSU}{Skobeltsyn Nuclear Physics Institute, 119899 Moscow, Russia}
\newcommand*{\NOWORSAY}{Institut de Physique Nucl\'eaire ORSAY, Orsay, France}
\newcommand*{\NOWINFNGE}{INFN, Sezione di Genova, 16146 Genova, Italy}

\author {H.Y.~Lu} 
\altaffiliation[Current address:]{\NOWIOWA}
\affiliation{\CMU}
\author {R.A.~Schumacher} 
\email[Contact: ]{schumacher@cmu.edu}
\affiliation{\CMU}
\author {K.P. ~Adhikari} 
\affiliation{\ODU}
\author {D.~Adikaram} 
\affiliation{\ODU}
\author {M.~Aghasyan} 
\affiliation{\INFNFR}
\author {M.J.~Amaryan} 
\affiliation{\ODU}
\author {S. ~Anefalos~Pereira} 
\affiliation{\INFNFR}
\author {J.~Ball} 
\affiliation{\SACLAY}
\author {M.~Battaglieri} 
\affiliation{\INFNGE}
\author {V.~Batourine} 
\affiliation{\JLAB}
% ...

\collaboration{CLAS Collaboration}
\noaffiliation

%
%
%
%
%
%


\date{\today}


\begin{abstract}
We report the first observation of the line shape of the
$\Lambda(1405)$ from electroproduction, and show that it is not a
simple Breit-Wigner resonance. Electroproduction of $K^+
\Lambda(1405)$ off the proton was studied by using data from CLAS at
Jefferson Lab in the range $1.0<Q^2<3.0$~(GeV/c)$^2$. The analysis
utilized the decay channels $\Sigma^+ \pi^-$ of the $\Lambda(1405)$
and $p \pi^0$ of the $\Sigma^+$. Neither the standard (PDG) resonance
parameters, nor free parameters fitting to a single Breit-Wigner
resonance represent the line shape. In our fits, the line shape
corresponds approximately to predictions of a two-pole meson-baryon
picture of the $\Lambda(1405)$, with a lower mass pole near
1368~MeV/c$^2$ and a higher mass pole near
1423~MeV/c$^2$. Furthermore, with increasing photon virtuality the
mass distribution shifts toward the higher mass pole.
\end{abstract}

\maketitle

\section{introduction}
The continued study of excited baryons is needed to refine our
knowledge of QCD. Probing states with peculiar properties is of
particular interest toward this goal. One such object is the
$\Lambda(1405)$, which has the quantum numbers
$I(J^P)=0(\frac{1}{2}^-)$, mass $m=1405.1$~MeV/c$^2$, and full width
$\Gamma = 50$~MeV/c$^2$ according to the Particle Data Group (PDG)
\citep{Beringer:1900zz}. It was first predicted by
\citet{Dalitz1959,Dalitz1960} as a quasi-bound $N\bar{K}$ state that
decays to $\Sigma\pi$, and was first observed by \citet{Alston1961} in
1961.

A quark model for the spectrum of baryons with a non-relativistic
harmonic oscillator potential and spin-spin hyperfine interaction
between the quarks was proposed by
\citet{Isgur1977,Isgur1978,Isgur1979}. This model reproduced all the
observed $P$-wave negative parity resonances well, except for the
$\Lambda(1405)$. The mass of the $\Lambda(1405)$ was predicted around
1490~MeV/c$^2$, leaving a discrepancy of about 80~MeV/c$^2$. The
model was extended to include relativistic dynamics by
\citet{Capstick1986} but this mass discrepancy remained. Following an
earlier paper by \citet{Oller2001}, \citet{Jido2003} used a chiral
unitary formalism to describe the $\Lambda(1405)$ as a composite of
two dynamically generated $I=0$ poles, at $m_0^{low} =
1.390+i0.066$~GeV/c$^2$ and $m_0^{high} =
1.420+i0.016$~GeV/c$^2$. Hence the key conclusion of this approach is
that the mass distribution, or ``line shape'' of the $\Lambda(1405)$
is formed by two poles in the complex energy plane instead of one
resonance. The higher mass pole couples more strongly to $N\bar{K}$,
while the lower mass pole couples more strongly to $\Sigma\pi$.
Furthermore, the line shape is expected to depend upon which reaction
channel excites it. For a recent review of the extensive literature on
baryonic systems within the chiral unitary framework, see
Ref.~\cite{Hyodo201255}. There are several other detailed predictions
for the parameters that characterize the poles making up the
$\Lambda(1405)$ mass region, for example
Refs.~\cite{Ikeda:2011pi,Ikeda:2012au,Guo:2012vv,Mai:2012dt,
Khemchandani:2011mf}.  One finds that theory is not in agreement 
about where these poles sit.

The most extensive experimental study of the $\Lambda(1405)$ mass
region used real photoproduction with data from
CLAS/JLab~\cite{Moriya,Schumacher2013} and also
LEPS/SPring-8~\cite{Ahn, Niiyama}.  Support for a strongly distorted
line shape was found, possibly consistent with two $I=0$ poles, as
well as evidence for interference with $I=1$ components of the
reaction mechanism. Further recent results came from $pp$ collisions
at HADES/GSI~\cite{Hades} and COSY/J\"ulich~\cite{Zychor}.
Low-statistics bubble-chamber hadronic beam data exist for
$K^-p$~\cite{Hemingway}, $K^-D$~\cite{Braun1977}, and
$\pi^-p$~\cite{Thomas} reactions.  Each case seems to show variations
in the mass distribution of the $\Lambda(1405)$.  To date there is no
theoretical work giving any prediction for electroproduction of the
$\Lambda(1405)$. The line shape seen in electroproduction could reveal
features not seen in other production channels. The virtual photon
exchanged in electroproduction ($Q^2 > 0$) may couple differently to
the $\Lambda(1405)$ than the real photon ($Q^2=0$). Therefore the
present study represents an exploration of an unknown corner of
$\Lambda(1405)$ phenomenology.

\begin{figure}
  \resizebox{5cm}{!}{\begin{fmffile}{electro} \fmfframe(1,7)(1,7){
    \begin{fmfgraph*}(210,162) \fmfleft{i2,i1}
    \fmfright{o5,o4,o3,o2,o1} \fmflabel{\Large $e$}{i1}
    \fmflabel{\Large $p$}{i2} \fmflabel{\Large $e^\prime$}{o1}
    \fmflabel{\Large $K^+$}{o2} \fmflabel{\Large $\pi^{-}$}{o3}
    \fmflabel{\Large $p$}{o4} \fmflabel{\Large $\pi^0$}{o5}
    \fmf{fermion}{i1,v1,o1} \fmf{fermion}{v2,o2} \fmf{fermion}{i2,v3}
    \fmf{fermion}{v4,o3} \fmf{fermion}{v5,o4} \fmf{fermion}{v5,o5}
    \fmf{photon,label={\Large $Q^{2}$}}{v1,v2}
    \fmf{plain,label={\Large $-t$}}{v2,v3} \fmf{double,label={\large
    $\Lambda(1405)$},l.side=left}{v3,v4} \fmf{double,label={\large
    $\Sigma^+$},l.side=right}{v4,v5} \fmfdot{v3,v4,v5}
    \fmfforce{0.3w,0.9h}{v1} \fmfforce{0.4w,0.6h}{v2}
    \fmfforce{0.4w,0.3h}{v3} \fmfforce{0.65w,0.3h}{v4}
    \fmfforce{0.8w,0.2h}{v5} \end{fmfgraph*} } \end{fmffile}   }
    \caption{\label{fig:feyn_diag}The presumed primary diagram for electroproduction of the
    $\Lambda(1405)$ at the kinematics of the present
    experiment.}
\end{figure}
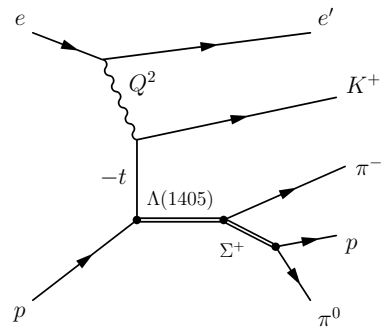

Figure~\ref{fig:feyn_diag} shows the presumed main contribution to the
$\Lambda(1405)$ electroproduction. Normally, $K$-nucleon scattering
cannot produce this state since it lies below threshold. However, the
off-shell nature of the exchanged strange meson allows the reaction to
proceed. The electron scattering process is characterized by the
four-momentum transfer $Q^2$, but this is one step removed from the
four-momentum of the exchanged kaon, $t$. At our kinematics $-t$
covers the range from 0.5 to 4.5~(GeV/c)$^2$.

This work analyzed data from the ``e1f'' run in Hall B at
Jefferson Lab collected using the CLAS spectrometer~\cite{Mecking2003}.  
It utilized a 5.5~GeV electron beam and a liquid hydrogen target.  The
$\Lambda(1405)$ decays solely to $\Sigma\pi$ according to the PDG
\citep{Beringer:1900zz}. The decay channel used in this analysis was
$\Sigma^+\pi^-$ for the $\Lambda$(1405) and $p\pi^0$ for the
$\Sigma^+$. Therefore, the final detected particles were the scattered
electron $e^-$ and the $K^+$, $p$, and $\pi^-$, while a $\pi^0$ was
missing for the exclusive electroproduction of the
$\Lambda(1405)$. Thus, events with exactly these four charged
particles were selected.

The data analysis is described in Sec.~\ref{sec:analysis}, followed by
presentation of the results in Sec.~\ref{sec:results}.  The results
are discussed and summarized in Sec.~\ref{sec:conclusions}.  Note that
a preliminary version of this analysis was published
previously~\cite{Lu2012}.

\section{data analysis}
\label{sec:analysis}

\begin{figure}
  \includegraphics[width=0.45\textwidth]{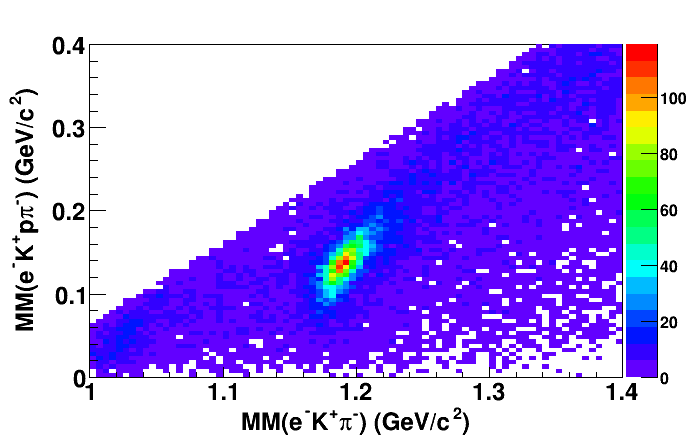}
  \caption{\label{fig:PizVSSigma}(Color online) Missing mass of $e^-K^+p\pi^-$ versus
  missing mass of $e^-K^+\pi^-$ from data. The strong central peak
  contains the $\Sigma^+\pi^0$ events selected for
  analysis.}
\end{figure}

\begin{figure}
  \subfloat{\label{fig:Yield_lowQ2}\includegraphics[width=0.45\textwidth]{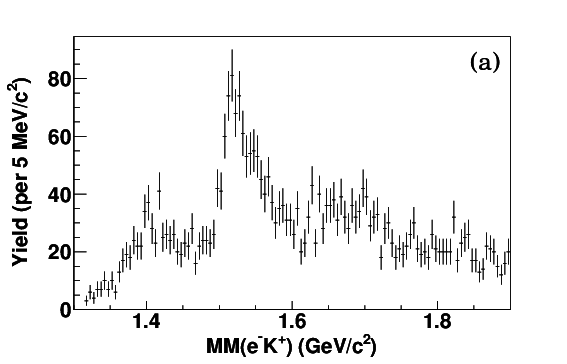}}

  \subfloat{\label{fig:Yield_highQ2}\includegraphics[width=0.45\textwidth]{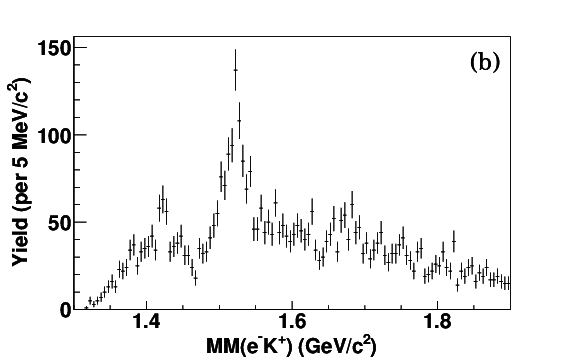}}
  \caption{\label{fig:rawYield}Number of counts plotted versus the
  missing mass of $e^-K^+$ without either acceptance corrections or
  background subtraction for (a) $1.0 \leq Q^2 \leq 1.5$~(GeV/c)$^2$,
  and (b) $1.5 \leq Q^2 \leq 3.0$~(GeV/c)$^2$.}
\end{figure}

The missing mass of $e^-K^+p\pi^-$ versus the missing mass of
$e^-K^+\pi^-$ is shown in Fig.~\ref{fig:PizVSSigma}. The missing
$\pi^0$'s and $\Sigma^+$'s are clearly seen near the center of
Fig.~\ref{fig:PizVSSigma}. A selection was made by requiring the
missing mass of $e^-K^+p\pi^-$ between 0.05 and 0.25~GeV/c$^2$ and the
missing mass of $e^-K^+\pi^-$ to lie between 1.140 and
1.240~GeV/c$^2$.  The $Q^2$ range of significance, after the
selection, was from 1.0 to 3.0~(GeV/c)$^2$. The $\Sigma^+\pi^-$
hyperon mass spectrum that results from computing the missing mass
from the $e^-K^+$ pair is shown in Fig.~\ref{fig:rawYield}. Two ranges
of $Q^2$ are shown, from 1.0 to 1.5~(GeV/c)$^2$ in
Fig.~\ref{fig:rawYield}\subref{fig:Yield_lowQ2}, and from 1.5 to
3.0~(GeV/c)$^2$ in
Fig.~\ref{fig:rawYield}\subref{fig:Yield_highQ2}. No acceptance
correction or background subtraction have been made at this stage. A
bump in the $\Lambda(1405)$ region and the $\Lambda(1520)$ peak are
clearly visible.

\begin{figure}
  \includegraphics[width=0.45\textwidth]{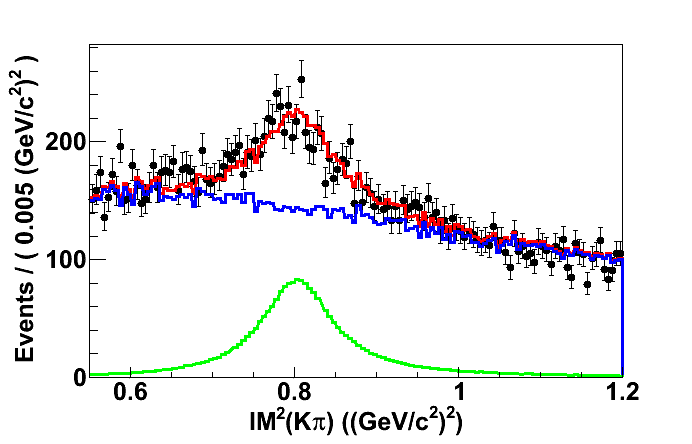} 
  \caption{\label{fig:FitRatio}(Color online) Invariant mass squared of $K^+\pi^-$. Points
  with error bars are measured data, the matching (red) line is the
  total fit to the data with two background simulations. The (green)
  line at the bottom is the simulation of $K^{*0}\Sigma^+$, and the
  upper (blue) line is simulation of $K^+\Sigma^+\pi^-$.
  }
\end{figure}

The background is mostly from non-resonant electroproduction of
$K^+\Sigma^+\pi^-$ and resonant $K^{*0}\Sigma^+$, where the $K^{*0}$
decays into $K^+$ and $\pi^-$. These background reaction channels were
simulated using the CLAS standard simulation package GSIM based on
GEANT \cite{GEANT1993}. Events were generated according to a phase
space distribution. The invariant mass of $K^+$ and $\pi^-$ was fitted
with simulations of the two background channels. The total fit to the
data set, with contributions from $K^{*0}\Sigma^+$ and
$K^+\Sigma^+\pi^-$, is shown in Fig.~\ref{fig:FitRatio}. The $K^{*0}$
background contribution to the combined simulated background channels
was determined from this fit. This fixed ratio was used in subsequent
fits to the line shapes in the hyperon mass region of interest.
 
\begin{figure}
  \includegraphics[width=0.45\textwidth]{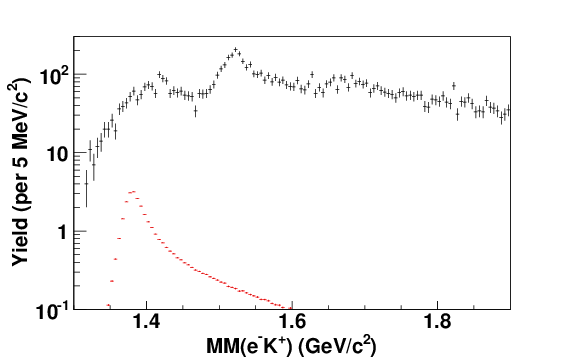}
  \caption{\label{fig:sigma_conta}(Color online) Estimated rescaled yield of the
  $\Sigma^{*0}$ contamination shown in red (lower points) and missing
  mass of $e^-K^+$ in black (upper points). The upper points are the
  sum of the yields from Fig.~\ref{fig:rawYield}. The histogram is on
  a log scale to make the $\Sigma^{*0}$ contribution
  visible.}
\end{figure}

A third production channel leading to the same final state particles
is electroproduction of $\Sigma^{*0}(1385)$ decaying to
$\Sigma^+\pi^-$. This contamination was measured by extracting the
yield of the $\Sigma^{*0}(1385)$ decaying through its dominant
$\Lambda\pi^0$ channel (branching fraction 88\%). This yield was
rescaled to $\Sigma^+\pi^-$ decay (6\%) and modified by the computed
respective acceptances. The contamination from $\Sigma^{*0}$(1385) was
thereby measured to be smaller than the statistical fluctuation in
each bin and therefore negligible. The estimated yield of
$\Sigma^{*0}(1385)$ and the measured yield of $\Sigma^+\pi^-$ are
shown in Fig.~\ref{fig:sigma_conta}. The histogram has a logarithmic
scale to make the estimated yield visible.

\begin{figure}
  \subfloat{\label{fig:accYield_lowQ2}\includegraphics[width=0.45\textwidth]{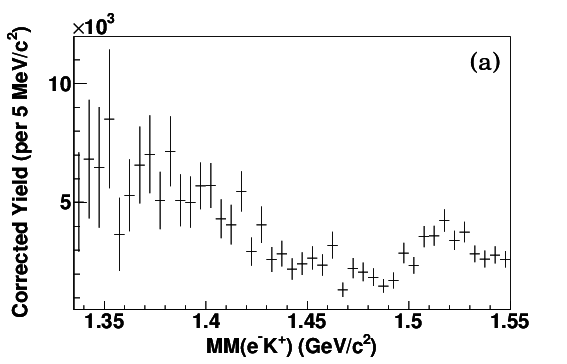}}

  \subfloat{\label{fig:accYield_highQ2}\includegraphics[width=0.45\textwidth]{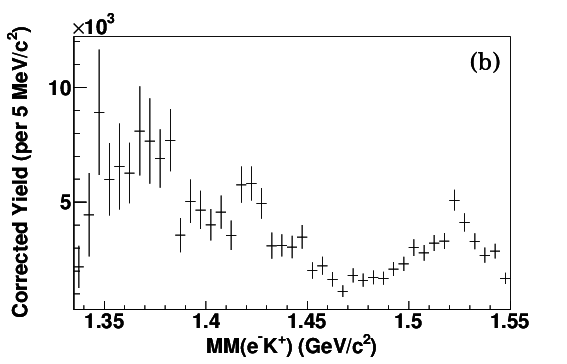}}
  \caption{\label{fig:accYield}Acceptance-corrected yield versus the
 missing mass of $e^-K^+$ for (a) $1.0 \leq Q^2 \leq 1.5$, and (b) $1.5 \leq Q^2 \leq 3.0$ (GeV/c)$^2$.}
\end{figure}

The acceptance correction was performed using the simulation of
non-resonant three-body phase space $K^+\Sigma^+\pi^-$. The generated
and accepted events were binned in a four dimensional space of the
independent kinematic variables, namely $Q^2$ (1.0 to
3.0~(GeV/c)$^2$), $W$ (1.5 to 3.5~GeV), the missing mass of $e^-K^+$,
and cosine of the kaon angle in the center-of-mass frame of
$K^+\Sigma^+\pi^-$. Another kinematic variable $\phi$, which is the
angle between the production plane characterized by $K^+$ and the
scattering plane characterized by the scattered $e^-$, was not
binned. This amounts to the assumption that $\sigma_{LT}$ and
$\sigma_{TT}$, the longitudinal-transverse and transverse-transverse
interference structure functions, are small in this reaction. The
dependence on this $\phi$ distribution was studied and the systematic
uncertainties were obtained as follows. The phase-space generated
Monte-Carlo events already match the measured data distributions in
$Q^2$, $W$, and particle momenta well. Furthermore, an improved
matching to data was achieved by selecting events in $e^-K^+$ missing
mass and $\phi$ to tailor the Monte-Carlo distributions used in
computing the acceptance. An acceptance factor was calculated bin by
bin in order to reduce bin-averaging effects.  The corrected hyperon
yield is shown in Fig.~\ref{fig:accYield}, where
Fig.~\ref{fig:accYield}\subref{fig:accYield_lowQ2} is for $Q^2$ from
1.0 to 1.5~(GeV/c)$^2$ and
Fig.~\ref{fig:accYield}\subref{fig:accYield_highQ2} for $Q^2$ from 1.5
to 3.0~(GeV/c)$^2$. The statistics in these two $Q^2$ regions are
comparable. The rightmost peak in both figures is the
$\Lambda(1520)$. A peak around 1.42~GeV/c$^2$ and another broader peak
at the lower invariant mass in
Fig.~\ref{fig:accYield}\subref{fig:accYield_highQ2} are not consistent
with the PDG values for the $\Lambda(1405)$
\cite{Beringer:1900zz}. However, neither peak is seen clearly in the
lower $Q^2$ range (Fig.~\ref{fig:accYield}\subref{fig:accYield_lowQ2}).

\section{results}
\label{sec:results}

\begin{figure}[!htbp]
  \subfloat{\label{fig:pdg_highQ2}\includegraphics[width=0.45\textwidth]{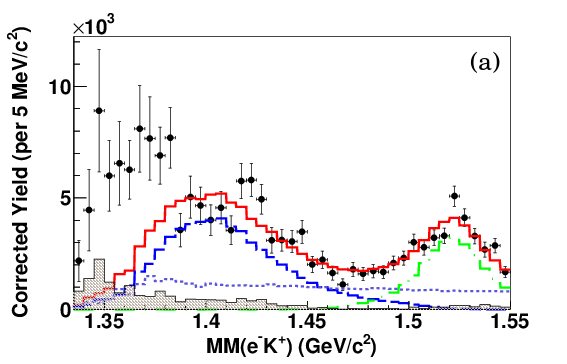}}

  \subfloat{\label{fig:onepole_highQ2}\includegraphics[width=0.45\textwidth]{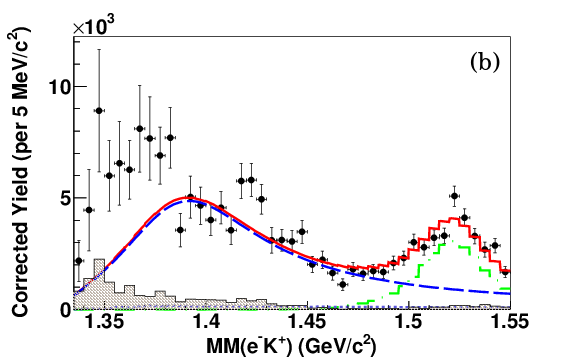}}

  \subfloat{\label{fig:twopole_highQ2}\includegraphics[width=0.45\textwidth]{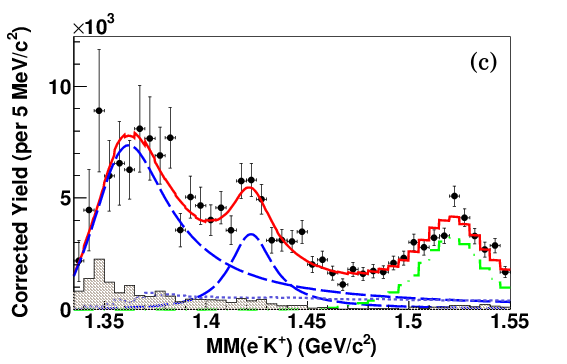}}
  \caption{\label{fig:fits_highQ2}(Color online) Fits of the missing
  mass of $e^-K^+$ for $1.5<Q^2<3.0$~(GeV/c)$^2$. Points with error
  bars are measured data, solid (red) lines are overall fits,
  dash-dotted (green) lines around 1.52~GeV/c$^2$ are from the
  $\Lambda(1520)$ simulation. The dashed (blue) lines are from the
  $\Lambda(1405)$ simulation parametrized by PDG values (a), by one
  relativistic Breit-Wigner function (b), and by two relativistic
  Breit-Wigner functions (c). The dotted (purple) lines show the
  summed background contributions. The shadowed histograms at the
  bottom show the estimated systematic uncertainty.}
\end{figure}

The higher $Q^2$ range distribution from
Fig.~\ref{fig:accYield}\subref{fig:accYield_highQ2} was fitted with
the simulated background, the $\Lambda(1520)$ and various assumptions
about the $\Lambda(1405)$. The results are shown in
Fig.~\ref{fig:fits_highQ2}. The black points with error bars are the
acceptance-corrected results. The shadowed histograms along the bottom
show the estimated systematic uncertainty in each mass bin.  The
systematic band was obtained by varying the selection of events (see
Fig.~\ref{fig:PizVSSigma}), the acceptance distribution in $\phi$, and
the acceptance in the hyperon invariant mass line shape (missing mass
of $e^-K^+$). Correlation between systematic uncertainties and
statistical fluctuations happens very often in low-statistics
experiments.  This appears clearly in the lower invariant mass
region. An empirical smoothing process to separate the fluctuations
from the estimated systematic uncertainties was implemented.
Quantitatively, the systematic uncertainties were obtained from the
following three analysis variations.

\begin{enumerate}

\item The range of the event selection was narrowed to between 
1.155 and 1.225~GeV/c$^2$ around the $\Sigma^+$ and to between 0.06
and 0.24~GeV/c$^2$ around the $\pi^0$. The resulting variations in
yields were smoothed by fitting the relative uncertainties to an
empirical smoothing function in the lower invariant mass region. The
resulting uncertainties were less correlated to the statistical
fluctuation and closer to the pure systematic uncertainties.  This
contributed about 45\% to the total bin-to-bin systematic uncertainty.

\item The phase-space Monte-Carlo distribution of events in $\phi$
matched the data fairly well. But an exact matching was enforced by
accepting Monte-Carlo events selectively. A similar smoothing process
was applied to the differences of acceptance-corrected yields between
a uniform phase-space distribution and the matching distribution.  The
resulting differences contributed about 35\% to the total systematic
uncertainty.

\item The phase-space Monte-Carlo line shape of the hyperon invariant 
mass was modified to match exactly the observed line shape. The
differences between the phase space distribution and the distribution
that was closely matched to measured data were smoothed by averaging
the neighboring bins. This systematic difference contributed about
20\% to the total bin-to-bin systematic uncertainty.

\end{enumerate}
The overall systematic bin-to-bin uncertainty shown in
Fig.~\ref{fig:fits_highQ2} was achieved by adding all pieces together
in quadrature. The relative changes in strength between the observed
peaks and the distortions of the peak shape due to radiative effects
are small in comparison to the precision of these results. Hence
radiative corrections were not performed.

In Fig.~\ref{fig:fits_highQ2} the resonances are computed as
incoherent relativistic Breit-Wigner functions of the form
\begin{equation}\label{eqn:rbw}
BW(m) \approx \frac{1}{2\pi}\frac{4mm_0\Gamma(q)}{(m^2-m^2_0)^2+(m_0\Gamma(q))^2}.
\end{equation}
In Eq.~\ref{eqn:rbw}, $\Gamma(q)=\frac{q}{q_0}\Gamma_0$, $m_0$, and $\Gamma_0$ are fit
parameters, $q$ is the momentum of $\pi^-$ in the hyperon rest frame,
and $q_0$ is value of the $q$ when $m=m_0$.

As seen in Fig.~\ref{fig:fits_highQ2}\subref{fig:pdg_highQ2}, a fit
using a single Breit-Wigner line shape with PDG parameters gives a
very poor representation of the data. The fit includes the summed
background contribution, simulation of $\Lambda(1520)$
electroproduction, and simulation of a Breit-Wigner ``resonance''
parametrized with the PDG values of the $\Lambda(1405)$. In
Fig.~\ref{fig:fits_highQ2}\subref{fig:onepole_highQ2}, an alternative
fit released these parameters from their PDG values. Also in this
case, the freely-fitted Breit-Wigner is a poor fit to the data. In
Fig.~\ref{fig:fits_highQ2}\subref{fig:twopole_highQ2}, we allowed the
$\Lambda(1405)$ to be represented as two incoherent Breit-Wigner
peaks. The best fit result gives the two peak positions at $m_0^{high}
= 1.423\pm0.002$~GeV/c$^2$ and $m_0^{low} = 1.368\pm0.004$~GeV/c$^2$,
where the uncertainties include only the fit errors. This fit is
clearly superior to the other two. The best-fit $\chi^2$ per degree of
freedom was 1.31
(Fig.~\ref{fig:fits_highQ2}\subref{fig:twopole_highQ2}) compared to
3.12 (Fig.~\ref{fig:fits_highQ2}\subref{fig:pdg_highQ2}) and 2.97
(Fig.~\ref{fig:fits_highQ2}\subref{fig:onepole_highQ2}),
respectively. A fit with two coherent Breit-Wigner line shapes was
also tested, resulting in a $\chi^2$ per degree of freedom 1.43,
similar to the best fit and qualitatively the same.

\begin{figure*}
  \subfloat{\label{fig:TwoResonance_10-30}\includegraphics[width=0.4\textwidth]{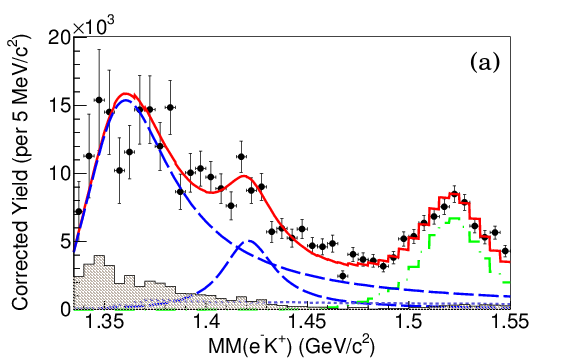}}
  \subfloat{\label{fig:TwoResonance_12-30}\includegraphics[width=0.4\textwidth]{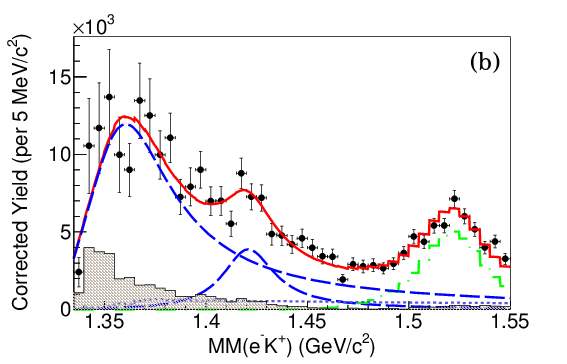}}

  \subfloat{\label{fig:TwoResonance_14-30}\includegraphics[width=0.4\textwidth]{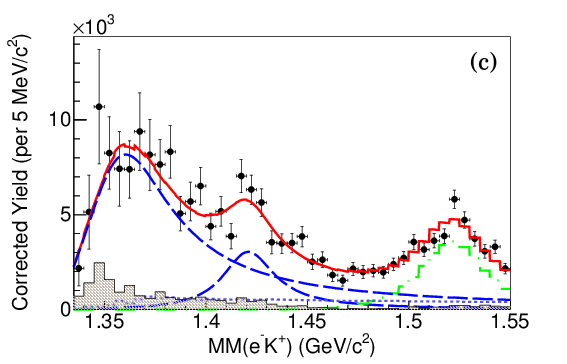}}
  \subfloat{\label{fig:TwoResonance_16-30}\includegraphics[width=0.4\textwidth]{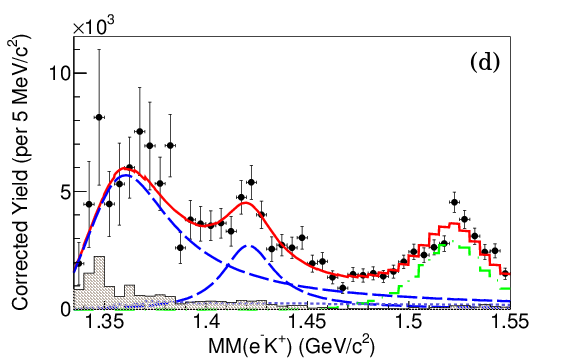}}

  \subfloat{\label{fig:TwoResonance_18-30}\includegraphics[width=0.4\textwidth]{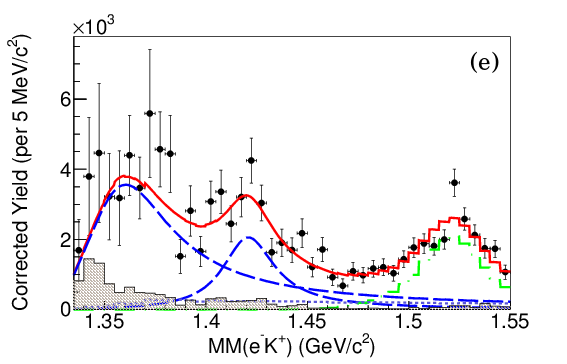}}
  \subfloat{\label{fig:TwoResonance_20-30}\includegraphics[width=0.4\textwidth]{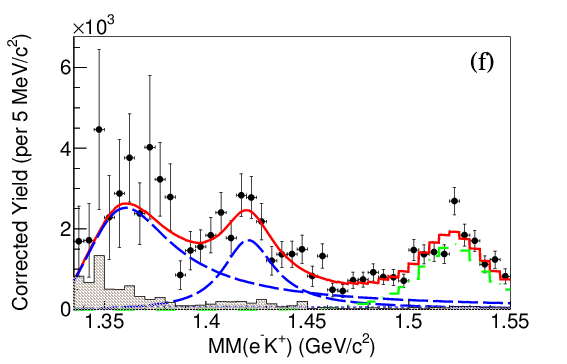}}

  \subfloat{\label{fig:TwoResonance_22-30}\includegraphics[width=0.4\textwidth]{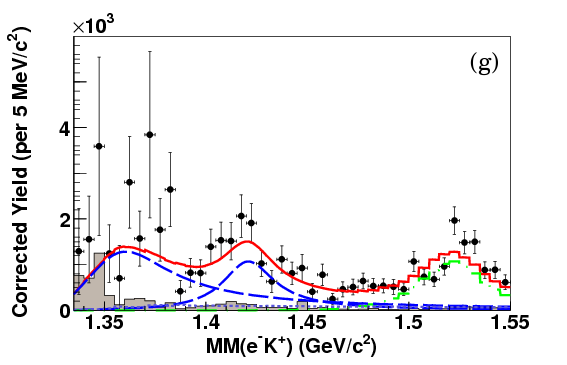}}

  \caption{(Color online) Overall fit of the acceptance-corrected
  missing mass of $e^-K^+$ with simulated background, simulated
  production of the $\Lambda(1520)$, and two relativistic Breit-Wigner
  functions in the ranges $Q^2_{min} \leq Q^2 \leq 3.0$~(GeV/c)$^2$,
  where $Q^2_{min}$ is: (a) 1.0~(GeV/c)$^2$, (b) 1.2~(GeV/c)$^2$, (c)
  1.4~(GeV/c)$^2$, (d) 1.6~(GeV/c)$^2$, (e) 1.8~(GeV/c)$^2$, (f)
  2.0~(GeV/c)$^2$, and (g) 2.2~(GeV/c)$^2$.  The fit takes the
  statistical uncertainties (error bars on points) into account. The
  shadowed histograms show the estimated systematic uncertainties. }
  \label{fig:TwoResonanceQ2}
\end{figure*}

Next, the acceptance-corrected yield was produced for seven $Q^2$
ranges, where the upper limit was fixed at 3.0~(GeV/c)$^2$ and the
lower limits were changed in steps from 1.0 to 2.2~(GeV/c)$^2$, as
shown in Fig.~\ref{fig:TwoResonanceQ2}.
Figure~\ref{fig:TwoResonanceQ2}\subref{fig:TwoResonance_10-30} shows
the data for $1.0 \leq Q^2 \leq 3.0$~(GeV/c)$^2$ and
Fig.~\ref{fig:TwoResonanceQ2}\subref{fig:TwoResonance_22-30} for $2.2
\leq Q^2 \leq 3.0$~(GeV/c)$^2$.  The histograms were fitted with two
relativistic incoherent Breit-Wigner functions, allowing only the
amplitudes to vary from range to range, while the centroids and widths
were common to all.  This combined fit extracted the two peak
positions $m_0^{high} = 1.422\pm0.002$~GeV/c$^2$, and $m_0^{low} =
1.365\pm0.002$~GeV/c$^2$, with a $\chi^2$ per degree of freedom of
1.44.  These values are consistent with the fit in
Fig.~\ref{fig:fits_highQ2}\subref{fig:twopole_highQ2}, though we note
that the error bars in this overall fit are correlated.  It can be
seen that the relative strength of the Breit-Wigner centered at
$m_0^{high}$ is increasing with increasing $Q^2$ compared to the one
centered at $m_0^{low}$, while staying comparable in strength to the
$\Lambda(1520)$.  This variation in relative strength with increasing
$Q^2$ is reminiscent of the expectation in chiral unitary models that
coupling to the two poles should depend on the coupling to the initial
state.  However, no model has attacked the question of whether such a
variation can be related to the $Q^2$ of a virtual photon.

The stability of these fits was tested by first adding and then
subtracting the systematic uncertainty (shadowed histograms) from the
corrected yields. The fits were repeated on the modified yield
distributions. The fit with two incoherent Breit-Wigner functions
still showed the best result in terms of reduced $\chi^2$. This shows
that the two-pole interpretation of our data remains the best, given
the estimated systematic uncertainties.

\section{discussion and conclusions}
\label{sec:conclusions}

The acceptance-corrected spectrum is qualitatively similar to an early
result from a bubble chamber experiment by \citet{Braun1977} for the
reaction $K^-D\rightarrow \Sigma^-\pi^+p$. \citet{Jido2009} proposed
the line shape as coming from interference of the $I=1$ $\Sigma^*(1385)$
and an $I=0$ pole at 1424~MeV/c$^2$. However, our electroproduction
work measured the amount of $\Sigma^{*0}(1385)$ and found negligible
contribution, as shown previously in Fig.~\ref{fig:sigma_conta}. The
expected yield of $\Sigma^{*0}(1385)$ is much smaller than the
statistical fluctuation of the observed statistics. In addition, any
interference between the $\Sigma^{*0}(1385)$ and the $\Lambda(1405)$
should vanish in this line shape. The reason is because the final line
shape stems from integration over all decay angles. The
$\Sigma^{*0}(1385)$ decays into $\Sigma \pi$ in $P$-wave, and the
$\Lambda(1405)$ decays in $S$-wave. The distributions are orthogonal
to each other and vanish after integrating over angles. Thus, we
conclude that the whole of the measured background-subtracted
distribution below the $\Lambda(1520)$ is due to the so-called
$\Lambda(1405)$.

It has been shown in photoproduction (at $Q^2=0$) that the structure
of the $\Lambda(1405)$ region is complex. Not only does the $N\bar K$
channel-coupling strongly distort the Breit-Wigner line shape used to
represent the $\Lambda(1405)$, but there is also a significant $I=1$
component to the reaction mechanism that is not coming from the
$\Sigma^{*0}(1385)$ \cite{Moriya,Schumacher2013}.  With the limited
statistical power of the present electroproduction data, it was not
feasible to analyze these data in a similar framework.  Higher
statistics data would clearly be useful here.

We note that the best-fit masses found in this analysis, $m_0^{low}$
and $m_0^{high}$, are close to a number of model predictions based on
unitarized coupled channels models.  Examples are the previously-quoted 
work of \citet{Jido2003}, and also 
\citet{Oller2001},
\citet{Ikeda:2011pi},
\cite{Ikeda:2012au},
\citet{Guo:2012vv}, and 
\citet{Khemchandani:2011mf}.  
Agreement is less good with the work of 
\citet{Mai:2012dt}
and the phenomenological photoproduction analysis of
\citet{Schumacher2013}.  Again, higher statistics electroproduction
data would clearly be useful.

To summarize, the $\Lambda(1405)$ mass region in electroproduction
studied in this work is poorly described by a single Breit-Wigner
distribution. Different fits have been tried and the best is with two
relativistic Breit-Wigner functions, which supports the two-pole model
as discussed in chiral unitary models~\cite{Hyodo201255}. In addition,
the relative strength of two line shapes changes with $Q^2$. The
progression suggests that at higher $Q^2$, which corresponds to more
off-shell exchange kaons (see Fig.~\ref{fig:feyn_diag}), it is more
likely to excite the higher mass pole of the $\Lambda(1405)$ structure
relative to the lower pole. The mass distribution seen in this
measurement also differs markedly from what is seen in photoproduction
at $Q^2=0$ \cite{Moriya,Schumacher2013}. The present limited
statistics makes more quantitative statements difficult, but the
existence of this phenomenon has been established and encourages
further exploration.

\begin{acknowledgments}
We acknowledge the outstanding efforts of the staff of the Accelerator
and Physics Divisions at Jefferson Lab that made this experiment
possible. The work of the Medium Energy Physics group at Carnegie
Mellon University was supported by DOE grant DE-FG02-87ER40315.  The
Southeastern Universities Research Association (SURA) operated the
Thomas Jefferson National Accelerator Facility for the United States
Department of Energy under contract DE-AC05-84ER40150.  Support was
also provided by the National Science Foundation, the United Kingdom's
Science and Technology Facilities Council (STFC), and the National
Research Foundation of Korea.
\end{acknowledgments}
\vfill




\bibliographystyle{apsrev4-1}   

\bibliography{CLAS_L1405_electroproduction}

\hyphenation{Post-Script Sprin-ger}
\begin{thebibliography}{29}%
\makeatletter
\providecommand \@ifxundefined [1]{%
 \@ifx{#1\undefined}
}%
\providecommand \@ifnum [1]{%
 \ifnum #1\expandafter \@firstoftwo
 \else \expandafter \@secondoftwo
 \fi
}%
\providecommand \@ifx [1]{%
 \ifx #1\expandafter \@firstoftwo
 \else \expandafter \@secondoftwo
 \fi
}%
\providecommand \natexlab [1]{#1}%
\providecommand \enquote  [1]{``#1''}%
\providecommand \bibnamefont  [1]{#1}%
\providecommand \bibfnamefont [1]{#1}%
\providecommand \citenamefont [1]{#1}%
\providecommand \href@noop [0]{\@secondoftwo}%
\providecommand \href [0]{\begingroup \@sanitize@url \@href}%
\providecommand \@href[1]{\@@startlink{#1}\@@href}%
\providecommand \@@href[1]{\endgroup#1\@@endlink}%
\providecommand \@sanitize@url [0]{\catcode `\\12\catcode `\$12\catcode
  `\&12\catcode `\#12\catcode `\^12\catcode `\_12\catcode `\%12\relax}%
\providecommand \@@startlink[1]{}%
\providecommand \@@endlink[0]{}%
\providecommand \url  [0]{\begingroup\@sanitize@url \@url }%
\providecommand \@url [1]{\endgroup\@href {#1}{\urlprefix }}%
\providecommand \urlprefix  [0]{URL }%
\providecommand \Eprint [0]{\href }%
\providecommand \doibase [0]{http://dx.doi.org/}%
\providecommand \selectlanguage [0]{\@gobble}%
\providecommand \bibinfo  [0]{\@secondoftwo}%
\providecommand \bibfield  [0]{\@secondoftwo}%
\providecommand \translation [1]{[#1]}%
\providecommand \BibitemOpen [0]{}%
\providecommand \bibitemStop [0]{}%
\providecommand \bibitemNoStop [0]{.\EOS\space}%
\providecommand \EOS [0]{\spacefactor3000\relax}%
\providecommand \BibitemShut  [1]{\csname bibitem#1\endcsname}%
\let\auto@bib@innerbib\@empty
\bibitem [{\citenamefont {Beringer}\ \emph {et~al.}(2012)\citenamefont
  {Beringer} \emph {et~al.}}]{Beringer:1900zz}%
  \BibitemOpen
  \bibfield  {author} {\bibinfo {author} {\bibfnamefont {J.}~\bibnamefont
  {Beringer}} \emph {et~al.} (\bibinfo {collaboration} {Particle Data Group}),\
  }\href {\doibase 10.1103/PhysRevD.86.010001} {\bibfield  {journal} {\bibinfo
  {journal} {Phys. Rev.}\ }\textbf {\bibinfo {volume} {D86}},\ \bibinfo {pages}
  {010001} (\bibinfo {year} {2012})}\BibitemShut {NoStop}%
\bibitem [{\citenamefont {Dalitz}\ and\ \citenamefont
  {Tuan}(1959)}]{Dalitz1959}%
  \BibitemOpen
  \bibfield  {author} {\bibinfo {author} {\bibfnamefont {R.~H.}\ \bibnamefont
  {Dalitz}}\ and\ \bibinfo {author} {\bibfnamefont {S.~F.}\ \bibnamefont
  {Tuan}},\ }\href {\doibase 10.1103/PhysRevLett.2.425} {\bibfield  {journal}
  {\bibinfo  {journal} {Phys. Rev. Lett.}\ }\textbf {\bibinfo {volume} {2}},\
  \bibinfo {pages} {425} (\bibinfo {year} {1959})}\BibitemShut {NoStop}%
\bibitem [{\citenamefont {Dalitz}\ and\ \citenamefont
  {Tuan}(1960)}]{Dalitz1960}%
  \BibitemOpen
  \bibfield  {author} {\bibinfo {author} {\bibfnamefont {R.}~\bibnamefont
  {Dalitz}}\ and\ \bibinfo {author} {\bibfnamefont {S.}~\bibnamefont {Tuan}},\
  }\href {\doibase 10.1016/0003-4916(60)90001-4} {\bibfield  {journal}
  {\bibinfo  {journal} {Annals of Physics}\ }\textbf {\bibinfo {volume} {10}},\
  \bibinfo {pages} {307 } (\bibinfo {year} {1960})}\BibitemShut {NoStop}%
\bibitem [{\citenamefont {Alston}\ \emph {et~al.}(1961)\citenamefont {Alston}
  \emph {et~al.}}]{Alston1961}%
  \BibitemOpen
  \bibfield  {author} {\bibinfo {author} {\bibfnamefont {M.}~\bibnamefont
  {Alston}} \emph {et~al.},\ }\href {\doibase 10.1103/PhysRevLett.6.698}
  {\bibfield  {journal} {\bibinfo  {journal} {Phys. Rev. Lett.}\ }\textbf
  {\bibinfo {volume} {6}},\ \bibinfo {pages} {698} (\bibinfo {year}
  {1961})}\BibitemShut {NoStop}%
\bibitem [{\citenamefont {Isgur}\ and\ \citenamefont {Karl}(1977)}]{Isgur1977}%
  \BibitemOpen
  \bibfield  {author} {\bibinfo {author} {\bibfnamefont {N.}~\bibnamefont
  {Isgur}}\ and\ \bibinfo {author} {\bibfnamefont {G.}~\bibnamefont {Karl}},\
  }\href {\doibase 10.1016/0370-2693(77)90074-0} {\bibfield  {journal}
  {\bibinfo  {journal} {Physics Letters B}\ }\textbf {\bibinfo {volume} {72}},\
  \bibinfo {pages} {109 } (\bibinfo {year} {1977})}\BibitemShut {NoStop}%
\bibitem [{\citenamefont {Isgur}\ and\ \citenamefont {Karl}(1978)}]{Isgur1978}%
  \BibitemOpen
  \bibfield  {author} {\bibinfo {author} {\bibfnamefont {N.}~\bibnamefont
  {Isgur}}\ and\ \bibinfo {author} {\bibfnamefont {G.}~\bibnamefont {Karl}},\
  }\href {\doibase 10.1103/PhysRevD.18.4187} {\bibfield  {journal} {\bibinfo
  {journal} {Phys. Rev. D}\ }\textbf {\bibinfo {volume} {18}},\ \bibinfo
  {pages} {4187} (\bibinfo {year} {1978})}\BibitemShut {NoStop}%
\bibitem [{\citenamefont {Isgur}\ and\ \citenamefont {Karl}(1979)}]{Isgur1979}%
  \BibitemOpen
  \bibfield  {author} {\bibinfo {author} {\bibfnamefont {N.}~\bibnamefont
  {Isgur}}\ and\ \bibinfo {author} {\bibfnamefont {G.}~\bibnamefont {Karl}},\
  }\href {\doibase 10.1103/PhysRevD.19.2653} {\bibfield  {journal} {\bibinfo
  {journal} {Phys. Rev. D}\ }\textbf {\bibinfo {volume} {19}},\ \bibinfo
  {pages} {2653} (\bibinfo {year} {1979})}\BibitemShut {NoStop}%
\bibitem [{\citenamefont {Capstick}\ and\ \citenamefont
  {Isgur}(1986)}]{Capstick1986}%
  \BibitemOpen
  \bibfield  {author} {\bibinfo {author} {\bibfnamefont {S.}~\bibnamefont
  {Capstick}}\ and\ \bibinfo {author} {\bibfnamefont {N.}~\bibnamefont
  {Isgur}},\ }\href {\doibase 10.1103/PhysRevD.34.2809} {\bibfield  {journal}
  {\bibinfo  {journal} {Phys. Rev. D}\ }\textbf {\bibinfo {volume} {34}},\
  \bibinfo {pages} {2809} (\bibinfo {year} {1986})}\BibitemShut {NoStop}%
\bibitem [{\citenamefont {Oller}\ and\ \citenamefont
  {Mei{\ss}ner}(2001)}]{Oller2001}%
  \BibitemOpen
  \bibfield  {author} {\bibinfo {author} {\bibfnamefont {J.}~\bibnamefont
  {Oller}}\ and\ \bibinfo {author} {\bibfnamefont {U.-G.}\ \bibnamefont
  {Mei{\ss}ner}},\ }\href {\doibase 10.1016/S0370-2693(01)00078-8} {\bibfield
  {journal} {\bibinfo  {journal} {Physics Letters B}\ }\textbf {\bibinfo
  {volume} {500}},\ \bibinfo {pages} {263 } (\bibinfo {year}
  {2001})}\BibitemShut {NoStop}%
\bibitem [{\citenamefont {Jido}\ \emph {et~al.}(2003)\citenamefont {Jido},
  \citenamefont {Oller}, \citenamefont {Oset}, \citenamefont {Ramos},\ and\
  \citenamefont {Mei{\ss}ner}}]{Jido2003}%
  \BibitemOpen
  \bibfield  {author} {\bibinfo {author} {\bibfnamefont {D.}~\bibnamefont
  {Jido}}, \bibinfo {author} {\bibfnamefont {J.}~\bibnamefont {Oller}},
  \bibinfo {author} {\bibfnamefont {E.}~\bibnamefont {Oset}}, \bibinfo {author}
  {\bibfnamefont {A.}~\bibnamefont {Ramos}}, \ and\ \bibinfo {author}
  {\bibfnamefont {U.-G.}\ \bibnamefont {Mei{\ss}ner}},\ }\href {\doibase
  10.1016/S0375-9474(03)01598-7} {\bibfield  {journal} {\bibinfo  {journal}
  {Nuclear Physics A}\ }\textbf {\bibinfo {volume} {725}},\ \bibinfo {pages}
  {181 } (\bibinfo {year} {2003})}\BibitemShut {NoStop}%
\bibitem [{\citenamefont {Hyodo}\ and\ \citenamefont
  {Jido}(2012)}]{Hyodo201255}%
  \BibitemOpen
  \bibfield  {author} {\bibinfo {author} {\bibfnamefont {T.}~\bibnamefont
  {Hyodo}}\ and\ \bibinfo {author} {\bibfnamefont {D.}~\bibnamefont {Jido}},\
  }\href {\doibase 10.1016/j.ppnp.2011.07.002} {\bibfield  {journal} {\bibinfo
  {journal} {Progress in Particle and Nuclear Physics}\ }\textbf {\bibinfo
  {volume} {67}},\ \bibinfo {pages} {55 } (\bibinfo {year} {2012})}\BibitemShut
  {NoStop}%
\bibitem [{\citenamefont {Ikeda}\ \emph {et~al.}(2011)\citenamefont {Ikeda},
  \citenamefont {Hyodo},\ and\ \citenamefont {Weise}}]{Ikeda:2011pi}%
  \BibitemOpen
  \bibfield  {author} {\bibinfo {author} {\bibfnamefont {Y.}~\bibnamefont
  {Ikeda}}, \bibinfo {author} {\bibfnamefont {T.}~\bibnamefont {Hyodo}}, \ and\
  \bibinfo {author} {\bibfnamefont {W.}~\bibnamefont {Weise}},\ }\href
  {\doibase 10.1016/j.physletb.2011.10.068} {\bibfield  {journal} {\bibinfo
  {journal} {Phys.Lett.}\ }\textbf {\bibinfo {volume} {B706}},\ \bibinfo
  {pages} {63} (\bibinfo {year} {2011})}\BibitemShut {NoStop}%
\bibitem [{\citenamefont {Ikeda}\ \emph {et~al.}(2012)\citenamefont {Ikeda},
  \citenamefont {Hyodo},\ and\ \citenamefont {Weise}}]{Ikeda:2012au}%
  \BibitemOpen
  \bibfield  {author} {\bibinfo {author} {\bibfnamefont {Y.}~\bibnamefont
  {Ikeda}}, \bibinfo {author} {\bibfnamefont {T.}~\bibnamefont {Hyodo}}, \ and\
  \bibinfo {author} {\bibfnamefont {W.}~\bibnamefont {Weise}},\ }\href
  {\doibase 10.1016/j.nuclphysa.2012.01.029} {\bibfield  {journal} {\bibinfo
  {journal} {Nucl.Phys.}\ }\textbf {\bibinfo {volume} {A881}},\ \bibinfo
  {pages} {98} (\bibinfo {year} {2012})}\BibitemShut {NoStop}%
\bibitem [{\citenamefont {Guo}\ and\ \citenamefont {Oller}(2013)}]{Guo:2012vv}%
  \BibitemOpen
  \bibfield  {author} {\bibinfo {author} {\bibfnamefont {Z.-H.}\ \bibnamefont
  {Guo}}\ and\ \bibinfo {author} {\bibfnamefont {J.}~\bibnamefont {Oller}},\
  }\href {\doibase 10.1103/PhysRevC.87.035202} {\bibfield  {journal} {\bibinfo
  {journal} {Phys.Rev.}\ }\textbf {\bibinfo {volume} {C87}},\ \bibinfo {pages}
  {035202} (\bibinfo {year} {2013})},\ \Eprint {http://arxiv.org/abs/1210.3485}
  {arXiv:1210.3485 [hep-ph]} \BibitemShut {NoStop}%
\bibitem [{\citenamefont {Mai}\ and\ \citenamefont
  {Meissner}(2013)}]{Mai:2012dt}%
  \BibitemOpen
  \bibfield  {author} {\bibinfo {author} {\bibfnamefont {M.}~\bibnamefont
  {Mai}}\ and\ \bibinfo {author} {\bibfnamefont {U.-G.}\ \bibnamefont
  {Meissner}},\ }\href {\doibase 10.1016/j.nuclphysa.2013.01.032} {\bibfield
  {journal} {\bibinfo  {journal} {Nucl.Phys.}\ }\textbf {\bibinfo {volume}
  {A900}},\ \bibinfo {pages} {51 } (\bibinfo {year} {2013})}\BibitemShut
  {NoStop}%
\bibitem [{\citenamefont {Khemchandani}\ \emph {et~al.}(2011)\citenamefont
  {Khemchandani}, \citenamefont {Martinez~Torres}, \citenamefont {Kaneko},
  \citenamefont {Nagahiro},\ and\ \citenamefont
  {Hosaka}}]{Khemchandani:2011mf}%
  \BibitemOpen
  \bibfield  {author} {\bibinfo {author} {\bibfnamefont {K.}~\bibnamefont
  {Khemchandani}}, \bibinfo {author} {\bibfnamefont {A.}~\bibnamefont
  {Martinez~Torres}}, \bibinfo {author} {\bibfnamefont {H.}~\bibnamefont
  {Kaneko}}, \bibinfo {author} {\bibfnamefont {H.}~\bibnamefont {Nagahiro}}, \
  and\ \bibinfo {author} {\bibfnamefont {A.}~\bibnamefont {Hosaka}},\ }\href
  {\doibase 10.1103/PhysRevD.84.094018} {\bibfield  {journal} {\bibinfo
  {journal} {Phys.Rev.}\ }\textbf {\bibinfo {volume} {D84}},\ \bibinfo {pages}
  {094018} (\bibinfo {year} {2011})}\BibitemShut {NoStop}%
\bibitem [{\citenamefont {Moriya}\ \emph {et~al.}(2013)\citenamefont {Moriya},
  \citenamefont {Schumacher} \emph {et~al.}}]{Moriya}%
  \BibitemOpen
  \bibfield  {author} {\bibinfo {author} {\bibfnamefont {K.}~\bibnamefont
  {Moriya}}, \bibinfo {author} {\bibfnamefont {R.~A.}\ \bibnamefont
  {Schumacher}},  \emph {et~al.} (\bibinfo {collaboration} {CLAS
  Collaboration}),\ }\href {\doibase 10.1103/PhysRevC.87.035206} {\bibfield
  {journal} {\bibinfo  {journal} {Phys. Rev. C}\ }\textbf {\bibinfo {volume}
  {87}},\ \bibinfo {pages} {035206} (\bibinfo {year} {2013})}\BibitemShut
  {NoStop}%
\bibitem [{\citenamefont {Schumacher}\ and\ \citenamefont
  {Moriya}(2013)}]{Schumacher2013}%
  \BibitemOpen
  \bibfield  {author} {\bibinfo {author} {\bibfnamefont {R.~A.}\ \bibnamefont
  {Schumacher}}\ and\ \bibinfo {author} {\bibfnamefont {K.}~\bibnamefont
  {Moriya}},\ }\href {\doibase 10.1016/j.nuclphysa.2013.03.003} {\bibfield
  {journal} {\bibinfo  {journal} {Nucl. Phys. A}\ } (\bibinfo {year} {2013}),\
  10.1016/j.nuclphysa.2013.03.003},\ \bibinfo {note} {(in press)}\BibitemShut
  {NoStop}%
\bibitem [{\citenamefont {Ahn}\ \emph {et~al.}(2003)\citenamefont {Ahn} \emph
  {et~al.}}]{Ahn}%
  \BibitemOpen
  \bibfield  {author} {\bibinfo {author} {\bibfnamefont {J.~K.}\ \bibnamefont
  {Ahn}} \emph {et~al.} (\bibinfo {collaboration} {LEPS Collaboration}),\
  }\href {\doibase 10.1016/S0375-9474(03)01164-3} {\bibfield  {journal}
  {\bibinfo  {journal} {Nucl. Phys.}\ }\textbf {\bibinfo {volume} {A721}},\
  \bibinfo {pages} {715} (\bibinfo {year} {2003})}\BibitemShut {NoStop}%
\bibitem [{\citenamefont {Niiyama}\ \emph {et~al.}(2008)\citenamefont {Niiyama}
  \emph {et~al.}}]{Niiyama}%
  \BibitemOpen
  \bibfield  {author} {\bibinfo {author} {\bibfnamefont {M.}~\bibnamefont
  {Niiyama}} \emph {et~al.} (\bibinfo {collaboration} {LEPS Collaboration}),\
  }\href {\doibase 10.1103/PhysRevC.78.035202} {\bibfield  {journal} {\bibinfo
  {journal} {Phys. Rev.}\ }\textbf {\bibinfo {volume} {C78}},\ \bibinfo {pages}
  {035202} (\bibinfo {year} {2008})}\BibitemShut {NoStop}%
\bibitem [{\citenamefont {Agakishiev}\ \emph {et~al.}(2012)\citenamefont
  {Agakishiev} \emph {et~al.}}]{Hades}%
  \BibitemOpen
  \bibfield  {author} {\bibinfo {author} {\bibfnamefont {G.}~\bibnamefont
  {Agakishiev}} \emph {et~al.} (\bibinfo {collaboration} {HADES
  Collaboration}),\ }\href {\doibase 10.1103/PhysRevC.85.035203} {\bibfield
  {journal} {\bibinfo  {journal} {Phys.Rev.}\ }\textbf {\bibinfo {volume}
  {C85}},\ \bibinfo {pages} {035203} (\bibinfo {year} {2012})}\BibitemShut
  {NoStop}%
\bibitem [{\citenamefont {Zychor}\ \emph {et~al.}(2008)\citenamefont {Zychor}
  \emph {et~al.}}]{Zychor}%
  \BibitemOpen
  \bibfield  {author} {\bibinfo {author} {\bibfnamefont {I.}~\bibnamefont
  {Zychor}} \emph {et~al.},\ }\href {\doibase 10.1016/j.physletb.2008.01.002}
  {\bibfield  {journal} {\bibinfo  {journal} {Phys. Lett.}\ }\textbf {\bibinfo
  {volume} {B660}},\ \bibinfo {pages} {167} (\bibinfo {year}
  {2008})}\BibitemShut {NoStop}%
\bibitem [{\citenamefont {Hemingway}(1985)}]{Hemingway}%
  \BibitemOpen
  \bibfield  {author} {\bibinfo {author} {\bibfnamefont {R.~J.}\ \bibnamefont
  {Hemingway}},\ }\href {\doibase 10.1016/0550-3213(85)90556-5} {\bibfield
  {journal} {\bibinfo  {journal} {Nucl. Phys.}\ }\textbf {\bibinfo {volume}
  {B253}},\ \bibinfo {pages} {742} (\bibinfo {year} {1985})}\BibitemShut
  {NoStop}%
\bibitem [{\citenamefont {Braun}\ \emph {et~al.}(1977)\citenamefont {Braun}
  \emph {et~al.}}]{Braun1977}%
  \BibitemOpen
  \bibfield  {author} {\bibinfo {author} {\bibfnamefont {O.}~\bibnamefont
  {Braun}} \emph {et~al.},\ }\href {\doibase 10.1016/0550-3213(77)90015-3}
  {\bibfield  {journal} {\bibinfo  {journal} {Nuclear Physics B}\ }\textbf
  {\bibinfo {volume} {129}},\ \bibinfo {pages} {1 } (\bibinfo {year}
  {1977})}\BibitemShut {NoStop}%
\bibitem [{\citenamefont {Thomas}\ \emph {et~al.}(1973)\citenamefont {Thomas},
  \citenamefont {Engler}, \citenamefont {Fisk},\ and\ \citenamefont
  {Kraemer}}]{Thomas}%
  \BibitemOpen
  \bibfield  {author} {\bibinfo {author} {\bibfnamefont {D.~W.}\ \bibnamefont
  {Thomas}}, \bibinfo {author} {\bibfnamefont {A.}~\bibnamefont {Engler}},
  \bibinfo {author} {\bibfnamefont {H.~E.}\ \bibnamefont {Fisk}}, \ and\
  \bibinfo {author} {\bibfnamefont {R.~W.}\ \bibnamefont {Kraemer}},\ }\href
  {\doibase 10.1016/0550-3213(73)90217-4} {\bibfield  {journal} {\bibinfo
  {journal} {Nucl. Phys.}\ }\textbf {\bibinfo {volume} {B56}},\ \bibinfo
  {pages} {15} (\bibinfo {year} {1973})}\BibitemShut {NoStop}%
\bibitem [{\citenamefont {Mecking}\ \emph {et~al.}(2003)\citenamefont {Mecking}
  \emph {et~al.}}]{Mecking2003}%
  \BibitemOpen
  \bibfield  {author} {\bibinfo {author} {\bibfnamefont {B.~A.}\ \bibnamefont
  {Mecking}} \emph {et~al.} (\bibinfo {collaboration} {CLAS Collaboration}),\
  }\href {\doibase 10.1016/S0168-9002(03)01001-5} {\bibfield  {journal}
  {\bibinfo  {journal} {Nucl. Inst. and Meth.}\ }\textbf {\bibinfo {volume}
  {503}},\ \bibinfo {pages} {513 } (\bibinfo {year} {2003})}\BibitemShut
  {NoStop}%
\bibitem [{\citenamefont {Lu}\ \emph {et~al.}(2012)\citenamefont {Lu},
  \citenamefont {Schumacher}, \citenamefont {Raue},\ and\ \citenamefont
  {Gabrielyan}}]{Lu2012}%
  \BibitemOpen
  \bibfield  {author} {\bibinfo {author} {\bibfnamefont {H.~Y.}\ \bibnamefont
  {Lu}}, \bibinfo {author} {\bibfnamefont {R.~A.}\ \bibnamefont {Schumacher}},
  \bibinfo {author} {\bibfnamefont {B.}~\bibnamefont {Raue}}, \ and\ \bibinfo
  {author} {\bibfnamefont {M.}~\bibnamefont {Gabrielyan}} (\bibinfo
  {collaboration} {CLAS Collaboration}),\ }\href {\doibase 10.1063/1.3701212}
  {\bibfield  {journal} {\bibinfo  {journal} {AIP Conf. Proc.}\ }\textbf
  {\bibinfo {volume} {1432}},\ \bibinfo {pages} {199} (\bibinfo {year}
  {2012})}\BibitemShut {NoStop}%
\bibitem [{\citenamefont {{CERN\_CN Division}}(1993)}]{GEANT1993}%
  \BibitemOpen
  \bibfield  {author} {\bibinfo {author} {\bibnamefont {{CERN\_CN Division}}},\
  }\href@noop {} {\enquote {\bibinfo {title} {{GEANT 3.2.1}},}\ }\bibinfo
  {howpublished} {{CERN Program Library W5013}} (\bibinfo {year}
  {1993})\BibitemShut {NoStop}%
\bibitem [{\citenamefont {Jido}\ \emph {et~al.}(2009)\citenamefont {Jido},
  \citenamefont {Oset},\ and\ \citenamefont {Sekihara}}]{Jido2009}%
  \BibitemOpen
  \bibfield  {author} {\bibinfo {author} {\bibfnamefont {D.}~\bibnamefont
  {Jido}}, \bibinfo {author} {\bibfnamefont {E.}~\bibnamefont {Oset}}, \ and\
  \bibinfo {author} {\bibfnamefont {T.}~\bibnamefont {Sekihara}},\ }\href
  {\doibase 10.1140/epja/i2009-10875-5} {\bibfield  {journal} {\bibinfo
  {journal} {Eu. Phys. J. A}\ }\textbf {\bibinfo {volume} {42}},\ \bibinfo
  {pages} {257} (\bibinfo {year} {2009})}\BibitemShut {NoStop}%
\end{thebibliography}%

\IfFileExists{\jobname.bbl}{}
 {\typeout{}
  \typeout{******************************************}
  \typeout{** Please run "bibtex \jobname" to obtain}
  \typeout{** the bibliography and then re-run LaTeX}
  \typeout{** twice to fix the references!}
  \typeout{******************************************}
  \typeout{}
 }

\end{document}